# Förster Resonance Energy Transfer and Laser Fluorescent Analysis of Defects in DNA Double Helix

**Vasil G. Bregadze[1], Zaza G. Melikishvili[2], Tamar G. Giorgadze[1], Zaza V. Jaliashvili[2], Jemal G. Chkhaberidze[1], Jamlet R. Monaselidze[1], Temur B. Khuskivadze[1]**

[1] *Department of Biological System Physics, Ivane Javakhishvili Tbilisi State University, Elevter Andronikashvili Institute of Physics, 6, Tamarashvili Str. 0177 Tbilisi, Georgia*

[2] *Department of Coherent Optics and Electronics, Georgian Technical University, Vladimir Chavchanidze Institute of Cybernetics, 5, Euli Str., 0186, Tbilisi, Georgia*

**Corresponding author:** Vasil G. Bregadze  v.bregadze@aiphysics.ge, vbregadze@gmail.com

**Keywords:** Laser induced fluorescence spectroscopy, Förster resonance energy transfer, DNA, acredine orange, ethidium bromide, high quality stability of double helix

## Abstract

Real time laser induced fluorescence spectroscopy usage for microanalysis of DNA double helix defects is shown. The method is based on Förster resonance energy transfer (FRET) in intercalator-donor pair (acridine orange as a donor and ethidium bromide as an acceptor). Transition metal ions such as Cu(II), Cu(I), Ag(I), silver nanoparticles (AgNPs), photo- and thermo effects were used to cause double helix defects in DNA.

FRET radii were experimentally estimated in background electrolyte solution (0.01 M $NaNO_3$) and proved to be $3.9 \pm 0.3$ nm and the data are in satisfactory agreement with the theoretically calculated value $R_0 = 3.5 \pm 0.3$ nm.

Concentration of DNA sites, exposed to Cu(II), Cu(I), Ag(I) ions, AgNPs impact as well as laser irradiation ($\lambda = 457$ nm) and temperature, which are applicable for intercalation, were estimated in relative units.

FRET method allows to estimate the concentration of double helix areas with high quality stability applicable for intercalation in DNA after it was subjected to stress effect. It gives the opportunity to compare DNAs of 1) different origin; 2) with various damage degrees; 3) being in various functional state.

**Introduction**

There is a tendency in modern medicine lately to use the so-called Förster resonance energy transfer (FRET) in donor-acceptor pair of dye molecules which intensely absorb light in the visible regions of spectrum and have significant quantum yield [1 - 6]. Microscopic FRET is also used for cytological diagnosis of tumors [7 - 14].

On the other hand, there are nearly no works where quantitative estimation of FRET applied to DNA is given. For instance, energy transfer (ET) effectiveness, quantitative estimation of DNA double helix conditions, accessibility of nuclear DNA to intercalator, etc. have not been yet estimated. In this connection it is up-to-day the modeling of DNA defects in solution and estimation of the quality of double helix. It is interesting to study a stress impact on the DNA by transitive metal (TM) ions, laser irradiation and heating.

Interaction of intercalator molecules with DNA, particularly, acridine orange (AO) and ethidium bromide (EB), depends on ionic strength of the solution, DNA nucleotide content, its sequence [15] and double helix structure. Besides, such interaction depends on transition metal ions (TM) [16] which cause ejection of intercalators, though intercalator molecules and TM ions have different binding sites on DNA. We should point out that AO and EB molecules as well as TM ions (Mn(II), Co(II), Ni(II), Cu(II) and Zn(II)) at interaction with DNA double helix have similar values of stability constants and their pK are in $4 - 6$ interval [15]. At interaction with DNA TM ions provoke point defects such as double proton transfer in GC pairs, depurinization, create inter strand cross-links [17] which are the reason of intercalators ejection from DNA [16]. In its turn, the above influences the efficiency of nonradiative transfer of electronic excitation energy from donor (D) to acceptor (A).

The base of Förster mechanism of nonradiative transfer of electronic excitation energy is the so-called inductive-resonance transfer of energy from D to A, where dipole-dipole interaction dominates. According to Förster in our case the rate of energy transfer $k_{ET}$ is in direct ratio to donor emission quantum yield ($q_{oD}$), overlap integral donor emission spectra and acceptor molecule extinction coefficient; and inversely to solvent refraction index in the fourth degree, distance between excited $D^*$ and A in the sixth degree, and $\tau_D$ is the donor emission decay time [18].

On the other hand, at inductive-resonance transfer of energy trough thin dielectric layers with alternative thickness d and permanent concentration of D and A the following correlation will be achieved [19, 20].

$$\frac{q_D}{q_{oD}} = 1 - \frac{q_A}{q_{oA}} = \left[1 + \left(\frac{d_0}{d}\right)^s\right]^{-1},$$ (1)

where s = 4, 6 and 2 are the energy transfers, consequently, for electric dipole − electric dipole, electric quadruple − electric dipole and magnetic dipole − electric dipole, $d_0$ − critical thickness of the layer. Correlation (1) in case of energy transfer from AO to EB intercalated in DNA double helix, can be re-written as follows:

$$\frac{q_D}{q_{oD}} = 1 - \frac{q_A}{q_{oA}} = \left[1 + \left(\frac{R_0}{R}\right)^s\right]^{-1},$$ (2)

where $q_{oD}$ is the quantum efficiency of donor fluorescence when the distance between energy donor and acceptor R→∞, $q_D$ at a given R, $q_{oA}$ is the quantum efficiency of sensibilized acceptor fluorescence at R→0 and $q_A$ at a given R The value $e_{ET} = 1 - q_{AO}/q_{oAO}$ was estimated as electron excitation energy transfer efficiency.

So, having definite concentration of intercalated pair D−A in DNA, and changing the concentration of DNA double helix the efficiency of energy transfer in the particular D-A pair can be sufficiently changed. Thus, the efficiency of energy transfer is proportional to DNA double helix site concentration. In Section 3 we show the validity of such approach for DNA-AO-EB ternary complexes.

The goal of the present research is development and application of laser induced fluorescence excitation energy transfer method to donor-acceptor intercalator pair for quantitative and qualitative study of stability quality DNA double helix in solution, in real time. The approach is based on the example of acridine orange molecule (donor) and ethidium bromide (acceptor) intercalated in DNA.

## 2. Materials and Methods

### 2.1. *Materials*

2.1.1. DNA. In our tests we used the Calf thymus DNA (40% GC), 'Sigma'. The concentration of nucleic acids was determined by UV absorption (spectrophotometer Specord M40, Carl Zeiss) using molar extinction coefficients ($\varepsilon$ = 6600 cm$^{-1}$ M$^{-1}$ at $\lambda$ = 260 nm). The double helix structure of the polymers was proved by their hyperchromicity (> 30%) and their typical thermal denaturation transition. (measured in 0.01 M NaNO$^3$, pH $\cong$ 6.0). pH was checked by pHmeter HANNA Instruments pH213.

2.1.2. Intercalators. Acridine orange (AO) was purchased from 'Sigma'. The concentration of the dye was determined colorimetrically at the isobestic point of the monomer-dimer system ($\lambda$=470 nm) using the molar extinction coefficients ($\varepsilon$ = 43 300 cm$^{-1}$M$^{-1}$). Ethidium bromide (EB) was also purchased from 'Sigma'. The concentration of the dye was determined colorimetrically ($\varepsilon$ = 5600 cm$^{-1}$M$^{-1}$ at $\lambda$ = 480 nm).

2.1.3. Ions. We used chemically pure copper chloride. Bidistillate water served as a solvent. In tests with Ag$^+$ ions chemically pure salts AgNO$_3$ were used and NaNO$_3$ served as background electrolytes.

2.1.4. Nanoparticles. Colloidal silver suspension with particle sizes of 1-2 nm in distilled water (200μg/ml) was purchased from DDS Inc., D/B/A, Amino Acid & Botanical Supply P.O Box 356, Cedar Knolls, NJ 07927.

### 2.2. *Instrumentation*

2.2.1. Absorption spectra of DNA complexes with intercalators AO and EB were registered in real time using CCD spectrometer AvaSpec ULS 2048-USB2. It should be underlined that registration of fluorescence  spectra excited by laser irradiation is necessary to carry out in real time as at such excitation of intercalators, AO in particular, its fast photo-oxidation takes place (see Fig. 1 and 2). Fig. 1 demonstrates laser irradiation effect ($\lambda$ = 457 nm) on absorption spectra of DNA complexes with AO. It is clearly seen that 20 min irradiation oxidizes 50% of AO molecules. Fig. 2 shows the data on laser irradiation effect on AO molecules in binary and ternary complexes AO – DNA, AO-DNA-Cu(II), AO- DNA-Cu(I) and AO-DNA-EB.

2.2.2. Diode laser SDL – 475 – 100T (Shanghai Dream Lasers Technology Co., Ltd.) was used for irradiation and excitation ($\lambda$ = 457 nm with optical beam cross-section 2 mm, and P = 200 mW) of laser induced fluorescence spectra.

2.2.3. Registration of fluorescence spectra of AO in binary and ternary DNA complexes was carried out in two ways:

2.2.3.1. In 1cm quartz fluorescent cells on luminescence spectrometer SDL-1 (Russia). The volume of samples under investigation was 1.5 ml. For fluorescence excitation of AO diode laser ($\lambda$ = 457 nm) was used. For real time fluorescence spectra registration (registration time 8 msec) CCD spectrometer was used. Fig. 3 demonstrates fluorescence spectra of binary and ternary AO complexes (AO-DNA and AO-DNA-EB). To the left of fluorescence spectrum Rayleigh scattering from laser radiation is clearly seen.

2.2.3.2. In glass rectangular cells of two sizes (height 6 and 12 mm, volume 40 and 80µl and diameter 3mm) intended for Raman scattering spectra registration with the application of light system DFS-24 (Russia) shown in Fig. 4. The light system was modified by our group to register fluorescence and is intended specially for experiments with laser excitation system. It is designed to provide the most effective application of excitation irradiation and irradiation of the light scattered by the sample. It is achieved by a) focusing the exciting irradiation in the sample volume; b) application of a high –aperture projection lenses and additional mirrors exercising multiple passage of the exciting beam through the sample, and the light, scattered in the opposite direction, is returned to the slit of the spectrometer. Fig. 5 presents fluorescence spectra of ternary AO-DNA-EB complexes. Just like in Fig. 3 to the left of fluorescence spectrum Relay scattered from laser light is clearly seen. The analysis of fluorescence spectra of ternary complex AO-DNA-EB shown in Fig. 3 and 5 presents good coincidence though the spectra were registered by various methods.

### 3. Results and Discussion

Fig.6 shows fluorescence spectra of binary and ternary AO-DNA and AO-ED-DNA complexes where concentration of DNA changes the distance between donor AO and acceptor EB. AO and EB concentrations were constant and equal to 0.14 x $10^{-4}$ M. DNA concentration varied from 0.5 x $10^{-4}$ to 10 x $10^4$ M per base pair (*bp*). Fig. 7 shows the ignition curves of AO fluorescence depending on the distance between AO and EB intercalated in DNA. The distance between donor and acceptor are given in nm and in *bp* units. The ignition curves of AO fluorescence are build in correspondence to Eq. (2) for s index equal to 4 , 6 or 2. An important characteristic of energy transfer process is Foerster distance $R_0$. At this distance, half of the donor molecules decay by energy transfer and the other half decays by the usual radiative and nonradiative rates [21].

$$R_0 = 0.211 \times 10^{-1} \left( \kappa^2 n^{-4} q_{oD} \, J(\lambda) \right)^{1/6} \text{ (in nm)} . \tag{3}$$

This expression allows the Förster distance to be calculated from the spectral properties ( $J(\lambda)$) of the donor and the acceptor and the donor quantum yield ($q_{oD}$), i.e. in terms of the experimentally known values taking into account the environment refractive index (n) and fluorescent molecules orientation factor ($\kappa^2$). In our case $J(\lambda) = 2.72 \times 10^{14}$ $M^{-1}(cm)^3(nm)^4$, $q_{oD} = 0.75$ [22], $\kappa^2 = 2/3$. For the index of refraction we choose n = 1.6 [23]. Thus, the Förster distance is calculated as $R_0 = 3.5 \pm 0.3$ which is in good agreement with experimental value $R_0 = 3.9 \pm 0.3$.

**Table 1.** Calculated values of $R_0$ for different *n* and $\kappa^2$

| *n* | $R_0$ (*bp*) | | $R_0$ (nm) | |
|---|---|---|---|---|
| | $\kappa^2 = \frac{2}{3}$ | $\kappa^2 = \frac{1}{2}$ | $\kappa^2 = \frac{2}{3}$ | $\kappa^2 = \frac{1}{2}$ |
| 1.33 | 11.72 | 11.17 | 3.98 | 3.79 |
| 1.40 | 11.32 | 10.79 | 3.85 | 3.67 |
| 1.60 | 10.36 | 9.87 | 3.52 | 3.35 |

Table 1 presents calculated values of $R_0$ for different $n$ and $\kappa^2$. As a refraction index some authors use $n = 1.33$ (water) [27], others assumed $n$ to be 1.40, which is valid for biomolecules in aqueous solution [21] or $n = 1.60$ in the case of DNA [23]. In all cases any dispersion effects are ignored. The term $\kappa^2$ is a factor describing the relative orientation of donor and acceptor transition dipoles in space. $\kappa^2$ is usually assumed to be equal to 2/3 which is appropriate for dynamic random average orientation of the donor and acceptor. . When the donor and acceptor are oriented in parallel planes, then $\kappa^2 = 1/2$ (for details see [21]). Evidently the orientation factor $\kappa^2$, and refractive index $n$ for different media do not affect significantly on $R_0$ value.

In Fig.7 we can see that theoretical curve describing electron excitation transfer for the case electric dipole – electric dipole (S = 4) corresponds to experimental data the best. Table 2 gives values for efficiency $e_{ET} = (1 - q_D/q_{oD})$ for AO intercalated in DNA depending on the distance between AO and EB given in *bp* units. The same table allows to estimate the distance between donor and acceptor in correspondence with the values of $e_{ET}$ evaluated from fluorescence spectra of FRET.

**Table2.** Efficiency of energy transfer $e_{ET} = (1 - q_D/q_{oD})$ from AO donor to EB acceptor depending on the distance R between them.

| $e_{ET}$ | R (*bp*) | $e_{ET}$ | R (*bp*) |
|---|---|---|---|
| 0.999 | 1 | 0.164 | 16 |
| 0.995 | 2 | 0.135 | 17 |
| 0.985 | 3 | 0.112 | 18 |
| 0.963 | 4 | 0.093 | 19 |
| 0.927 | 5 | 0.078 | 20 |
| 0.872 | 6 | 0.066 | 21 |
| 0.800 | 7 | 0.055 | 22 |
| 0.715 | 8 | 0.047 | 23 |
| 0.622 | 9 | 0.040 | 24 |
| 0.529 | 10 | 0.035 | 25 |
| 0.442 | 11 | 0.030 | 26 |
| 0.365 | 12 | 0.026 | 27 |
| 0.300 | 13 | 0.023 | 28 |
| 0.245 | 14 | 0.020 | 29 |
| 0.201 | 15 | 0.018 | 30 |

It is well known and it was shown in our work [17] too that transition metal ions at interaction with DNA cause or participate in different conformational changes, e.g., Cu(II) ions initiate DNA transition from B to C form [24]; with the help of TM ions and ethanol B-Z transition can be initiated [17]. Effect of Ag(I) ions on structure of DNA is known [25]. As far as in 1996 we discovered ejection of AO and EB intercalators from DNA. Besides, the rise of temperature in the solution causes melting of DNA double helix. In this connection it was interesting to investigate the electron excitation transfer in D-A pairs intercalated in DNA that is under the effect of different stress factors with the aim of finding intact sites of DNA double helix.

Figs. 8, 9, 10 and 11 demonstrate the influence of Cu(II), Cu(I), Ag(I), Ag NPs and laser irradiation ($\lambda$ = 457 nm) on electron excitation energy transfer efficiency from AO to EB intercalated in DNA which is shown in the rise of effectiveness FRET. The estimated data are given in Table 3. There are also the data for the distance between AO and EB in bp units, as well as the relative concentrations of DNA sites applicable for intercalation. The Figs. 8-11 also show that these ions quench AO fluorescence – the phenomena was studied by our group in [16, 26] and it is connected with doth dynamic quenching and the quenching caused by nonradiative transfer of excitation energy. On the one hand, Cu(II), Cu(I), Ag(I), Ag NPs and laser irradiation ($\lambda$ = 457 nm) quench AO fluorescence and on the other hand, increase FRET intensity. It obviously points out that there are different reasons for the phenomenon, in particular, FRET intensity is connected with the quality of double helix, i.e. stability constant of AO and EB with DNA.

Table 3. Cu(II), Cu(I), Ag(I) ions, AgNP, laser irradiation ($\lambda$= 457 nm) and heating effects on $e_{ET}$[1] and $C_{dh}^{st}/C_0 = R_{AO-EB}^{st}/R_{AO-EB}^0$[2]

| Stress factor for DNA - AO - EB [3] | $e_{ET}$ (%) | $R_{AO-EB}$ (bp) [4] | $\dfrac{C_{dh}^{st}}{C_0} = \dfrac{R_{AO-EB}^{st}}{R_{AO-EB}^0}$ |
|---|---|---|---|
| - | 20 | 15 | 1 |
| Cu(II) | 58 | 10-9 | 0.63 |
| Cu(II)+$h\nu$ | 77 | 8-7 | 0.5 |
| Cu(I) | 53 | 10 | 0.67 |
| Cu(I)+$h\nu$ | 81 | 7 | 0.47 |
| Ag(I) | 67 | 8-9 | 0.57 |
| AgNPs (1) | 62 | 9 | 0.6 |
| AgNPs (2) | 76 | 8-7 | 0.5 |
| **Heating** | | | |
| $20^0$C | 20 | 15 | 1 |
| $50^0$C | 48 | 10-11 | 0.7 |
| $60^0$C | 52 | 10 | 0.67 |
| $70^0$C | 76 | 8-7 | 0.5 |
| $80^0$C | 81 | 7 | 0.47 |
| $90^0$C | 82 | 7 | 0.47 |
| **Boiling** | | | |
| $20^0$C | 30 | 13 | 1 |
| $100^0$C, 5 min | 80 | 7 | 0.54 |
| $100^0$C, 10 min | 87 | 6 | 0.46 |
| $100^0$C, 20 min | 95 | 4 | 0.31 |

[1] Förster resonance electron excitation energy transfer from donor AO to acceptor EB. [2] relative concentration of DNA double helix areas applicable for AO and EB intercalation, where $C_{dh}^{st}$ is concentration of double helix areas in *bp* left after stress effect, $C_0$ – initial DNA concentration in M *bp*, $R_{AO-EB}^0$ – distance between AO and EB at initial DNA concentrations, $R_{AO-EB}^{st}$ – distance between AO and EB after stress; [3] different effects on DNA double helix; [4] $R_{AO-EB}$ – distance between AO and EB in *bp* units evaluated from efficiency $e_{ET}$ ( see Table 2).

Besides, temperature effect on FRET stability was investigated. Fig.12 shows heating effect on DNA solution located in a hermetic test-tube in thermostat. At various temperatures T = 50, 60, 70, 80 and 90 $^o$C, 2 ml of DNA solution was taken out and put into icy bath (T = 0 $^o$C), then at 20 $^o$C intercalator pair AO-EB was added to the solution and fluorescence spectra were registered.

Fig.13 shows fluorescence spectra for ternary complexes AO-DNA-EB. The complexes were prepared as follows: DNA solutions put in hermetic test-tubes were kept in vitro in thermal bath at the temperature $100^0$ C for different periods of time (5, 10 and 20 min). Then they were quickly cooled in icy bath (T = 0 $^o$C). After the procedure intercalator pair AO-EB was added to the solution and fluorescence spectra were registered. From the spectra given in Figs. 12, 13 and with the application of Table 2 the effectiveness of FRET, the data for the distance between AO and EB in bp units, as well as the relative concentrations of DNA sites applicable for intercalation were estimated. The results are given in Table 3.

Analyzing data given in Table 3 we can make a conclusion that the ions of the investigated transition metals, laser irradiation of AO and the effect of heating decrease the concentration of undamaged areas of DNA double helix, i.e. the sites able to intercalate dye molecules such as AO and EB. In its turn, it leads to increase of intercalator concentration per undamaged area of double helix, i.e. to decrease of the distance between AO and EB and finally to the increase of FRET effectiveness. Thus, the proposed laser-induced fluorescence method FRET allows to estimate the concentration of double helix areas with high quality stability applicable for intercalation in DNA after it was subjected to stress effect. It gives the opportunity to compare DNAs of 1) different origin; 2) with various damage degrees; 3) being in various functional state.

**Conclusions**

The method of laser induced fluorescence excitation energy transfer to donor-acceptor intercalator pair for quantitative and qualitative study of stability quality DNA double helix in solution, in real time, is offered. The approach is based on the example of acridine orange molecule (donor) and ethidium bromide (acceptor) intercalated in DNA.

It is shown that ions Cu(II), Cu(I), Ag(I) and Ag NPs, laser irradiation of AO and the effect of heating decrease the concentration of undamaged areas of DNA double helix, i.e. the sites able to intercalate dye molecules such as AO and EB.

FRET radii were experimentally estimated in background electrolyte solution (0.01 M NaNO$_3$) and proved to be 3.9 ± 0.3 nm and the data are in satisfactory agreement with the theoretically calculated value R$_0$ = 3.5 ± 0.3 nm.

FRET method allows to estimate the concentration of double helix areas with high quality stability applicable for intercalation in DNA after it was subjected to stress effect. It gives the opportunity to compare DNAs of 1) different origin; 2) with various damage degrees; 3) being in various functional state.

### Acknowledgements

The authors express their gratitude to Prof. *Henri Kagan,* Paris-Sud University, Orsay, for his permanent kind to our work attention and to Mrs. G. Nijaradze for her help in preparing the manuscript. The work was partly supported by the grant of Shota Rustaveli National Science Foundation number GNSF/ST09_508_2-230.

### References

[1]  X. Chen, B. Zehnbauer, A. Gnirke, and P.-Y. Kwok, "Fluorescence energy transfer detection as a homogeneous DNA diagnostic method," *Proc. Nat. Acad. Sci. U.S.A.*, **94**(20), 10756-10761 (1997).

[2]  R. B. Sekar and A. Periasamy, "Fluorescence resonance energy transfer (FRET) microscopy imaging of live cell protein localizations," *The Journal of Cell Biology*, **160**(5), 629-633 (2003).

[3]  C.-W. Chang, M. Wu, S. D. Merajver, and M.-A. Mycek, "Physiological fluorescence lifetime imaging microscopy improves Förster resonance energy transfer detection in living cells," *J. Biomed. Opt*. **14**(6), 060502 (2009).

[4]     S. C. Hovan, S. Howell and P. S.-H. Park, "Förster resonance energy transfer as a tool to study photoreceptor biology," *J. Biomed. Opt.* **15**(6), 067001 (2010).

[5]     Y. Wang and L. V. Wang, "Förster resonance energy transfer photoacoustic microscopy," *J. Biomed. Opt.* **17**(8), 086007 (2012).

[6]     A. P. Siegel, N. M. Hays, and R. N. Day, "Unraveling transcription factor interactions with heterochromatin protein 1 using fluorescence lifetime imaging microscopy and fluorescence correlation spectroscopy," *J. Biomed. Opt.* **18**(2), 025002 (2013).

[7]     S. Murata, P. Herman, H.-J. Lin, and J. R. Lakowicz, "Fluorescence Lifetime Imaging of Nuclear DNA: Effect of Fluorescence Resonance Energy Transfer," *Cytometry* **41**,178-185 (2000).

[8]     S. Murata, P. Herman, and J. R. Lakowicz, "Texture Analysis of Fluorescence Lifetime Images of AT- and GC-rich Regions in Nuclei," *J Histochem Cytochem* **49**, 1443-1451 (2001).

[9]     T. Nakazawa, T. Kondo, Y. Kobayashi, N. Takamura, S. Murata, K. Kameyama, A. Muramatsu, K. Ito, M. Kobayashi, and R. Katoh "RET Gene Rearrangements (RET/PTC1 and RET/PTC3) in Papillary Thyroid Carcinomas from an Iodine-Rich Country (Japan)," *CANCER* **104**(5), 943-951 (2005).

[10]    S. Murata, P. Herman, M. Iwashina, K. Mochizuki, T. Nakazawa, T. Kondo, J. R. Lakowicz, and R. Katoh, "Application of microscopic Förster resonance energy transfer to cytological diagnosis of the thyroid," *J. Biomed. Opt*. **10**(3), 034008 (2005).

[11]    K. Mochizuki, T. Kondo, T. Nakazawa, M. Iwashina, T. Kawasaki, N. Nakamura, T. Yamane, S. Murata, K. Ito, K. Kameyama, M. Kobayashi, and R. Katoth, "RET rearrangements and BRAF mutation in undifferentiated thyroid carcinomas having papillary carcinoma components," *Histopathology* **57**(3), 444-450 (2010).

[12]    F. Feng, L. Liu, and S. Wang, "Fluorescent conjugated polymer-based FRET technique for detection of DNA methylation of cancer cells," *Nature Protocols* **5**, 1237 - 1246 (2010).

[13]    S. Lu, and Y. Wang, "FRET Biosensors for Cancer Detection and Evaluation of Drug Efficacy," *Clin. Cancer. Res*., **16**(15), 3822–3824 (2010).

[14]    A. San Martín, S. Ceballo, I. Ruminot, R. Lerchundi, W.B. Frommer, L. F. Barros, "A Genetically Encoded FRET Lactate Sensor and Its Use To Detect the Warburg Effect in Single Cancer Cells," *PLoS ONE* **8**(2), e57712 (2013).

[15]    V.G. Bregadze, In: Metal Ions in Biological Systems. A. Sigel and H. Sigel, eds.  New York: Marcel Dekker, 1996, Vol.32 , 453-474.

[16]    V.G. Bregadze, J.G. Chkhaberidze, I.G. Khutsishvili, in: A. Sigel, H. Sigel (Eds.), Metal Ions in Biological Systems, vol. 33 (chapter 8), Marcel Dekker, New York, 1996, p. 269.

[17]    V. Bregadze, E. Gelagutashvili, K. Tsakadze, Thermodynamic Models Of Metal Ion-DNA Interactions; in: Metal-Complex DNA Interactions, N. Hajiliadis and E. Sletten (Eds), 31-53 (Blackwell Publishing, Oxford, UK), 2009.

[18]    J. A. Barltrop and J. D. Coyle, Excited States in Organic Chemistry, Wiley-lnterscience,. New York, 1977.

[19]    V.L. Ermolaev, E.N. Bodunov, E.B. Sveshnikova, T.A. Shakhverdov, in: M.D. Galanin (Ed.), Radiationless Electron Excitation Energy Transfer, Nauka, Leningrad, 1977.

[20]    H. Kuhn, "Classical Aspects of Energy Transfer in Molecular Systems," *J. Chem. Phys.* **53**, 101-108 (1970).

[21]    J. R. Lakowicz, "*Principles of Fluorescence Spectroscopy,*" Third Edition (Springer, New York, 1983).

[22]    R. E. Harrington, "Flow birefringence of persistence length deoxyribonucleic acid. Hydrodynamic properties, optical anisotropy, and hydration shell anistropy," *J Am Chem Soc.*, **92**(23), 6957-6964 (1970).

[23]  Y. Kubota, and R. F. Steiner, "Fluorescence decay and quantum yield characteristics of acridine orange and proflavine bound to DNA," *Biophys Chem.*, **6**(3), 279-289 (1977).

[24]  W. Förster, E. Bauer, H. Schütz, H. Berg, N. M. Akimenko, L. E. Minchenkova, Yu. M. Evdokimov, and Ya. M. Varshavsky, "Thermodynamics and kinetics of the interaction of copper (II) ions with native DNA," *Biopolymers* **18**(3), 625-661 (1979).

[25]  F. X. Wilhelm and M. Daune, "Interactions des ions métalliques avec le DNA. III. Stabilité et configuration des complexes Ag-DNA" *Biopolymers* **8**, 121-137 (1969).

[26]  V. G. Bregadze, I. G. Khutsishvili, J. G. Chkhaberidze, K. Sologashvili, "DNA as a mediator for proton, electron and energy transfer induced by metal ions," *Inorganica Chimica Acta*, 2002, V. 339, 145-159.

[27]  L. Navotny and B. Hecht, "*Principles of Nano-Optics*," (Cambridge Univesity Press, Cambridge, New York, 2006).

**Figures**

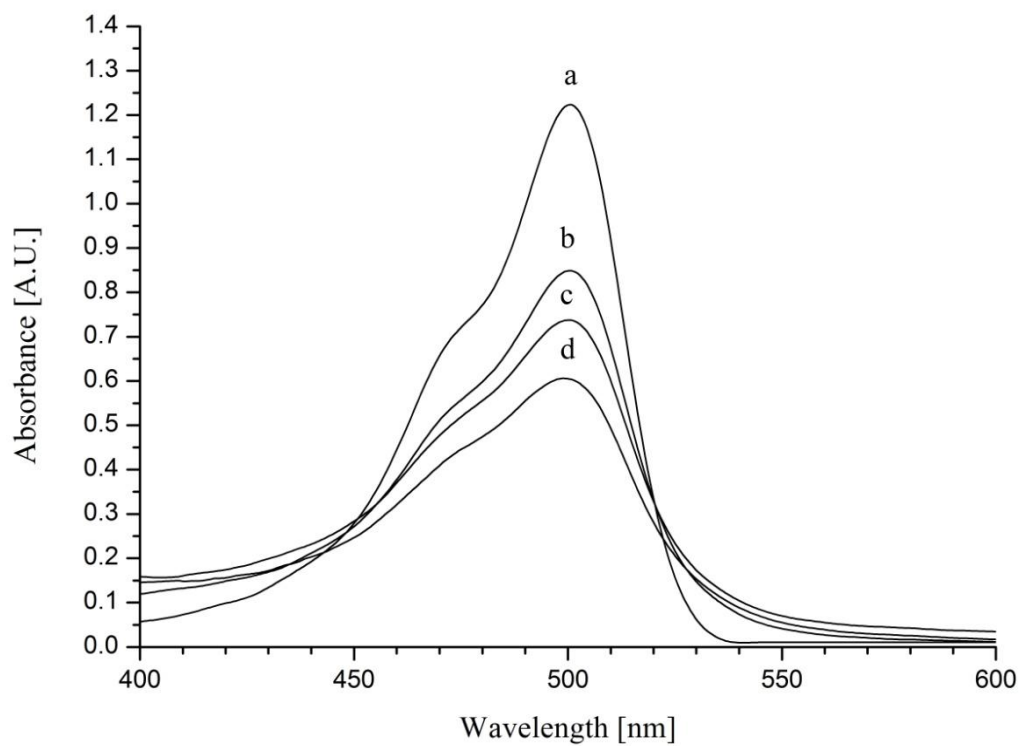

**Fig.1** Absorbtion spectra of DNA-AO before and after laser ($\lambda$= 457nm) irradiation: (a) DNA-AO (0); (b) DNA-AO (7 min); (c) DNA-AO (14 min); (d) DNA-AO (21min). [DNA]-$0.7 \cdot 10^{-3}$ M (P), [AO]-$0.7 \cdot 10^{-4}$ M, [NaNO$_3$]-$10^{-2}$ M.

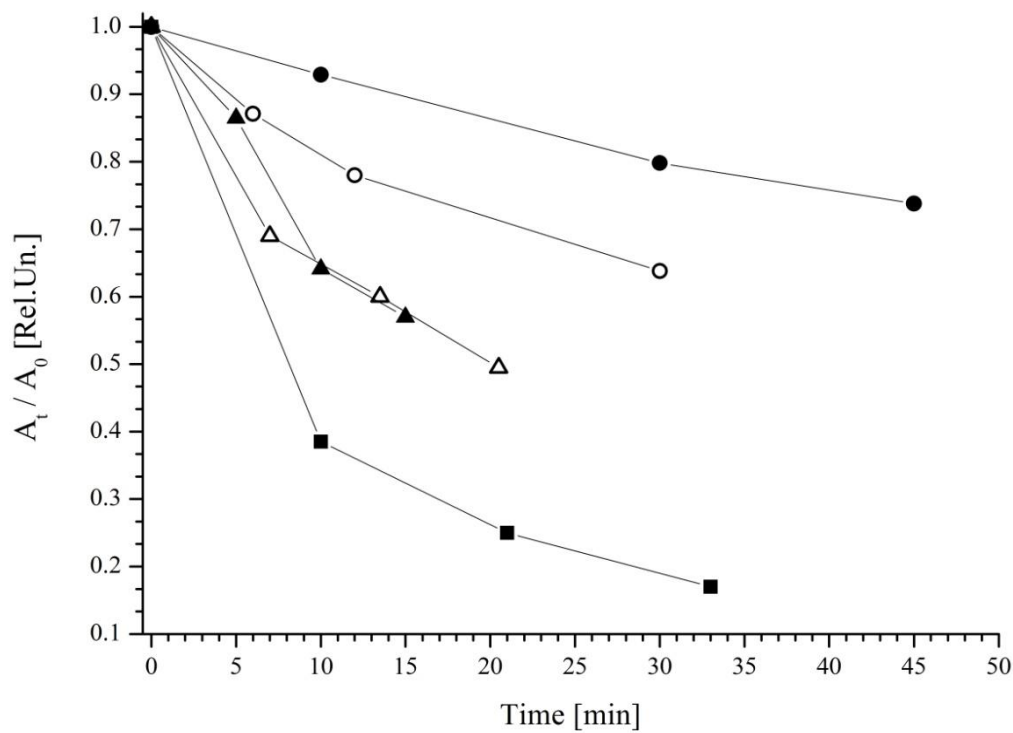

**Fig. 2** Laser irradiation effect on AO molecules in binary and ternary complexes AO-DNA, AO-DNA-Cu(II), AO- DNA-Cu(I) and AO-DNA-EB. ● — AO-DNA-Cu(I); ○ — AO-DNA-Cu(II), ▶ — AO-DNA-EB ▷ — AO-DNA, ■ — DNA-AO-Ag (I). [DNA]–$0.7 \cdot 10^{-3}$ M (P), [AO]–$0.7 \cdot 10^{-4}$ M, [EB]–$0.7 \cdot 10^{-4}$ M, [AA]–$1.4 \cdot 10^{-4}$ M, [$Ag^+$]–$0.7 \cdot 10^{-4}$ M, [$CuCl_2$]–$0.7 \cdot 10^{-4}$ M, [$NaNO_3$]–$10^{-2}$ M.

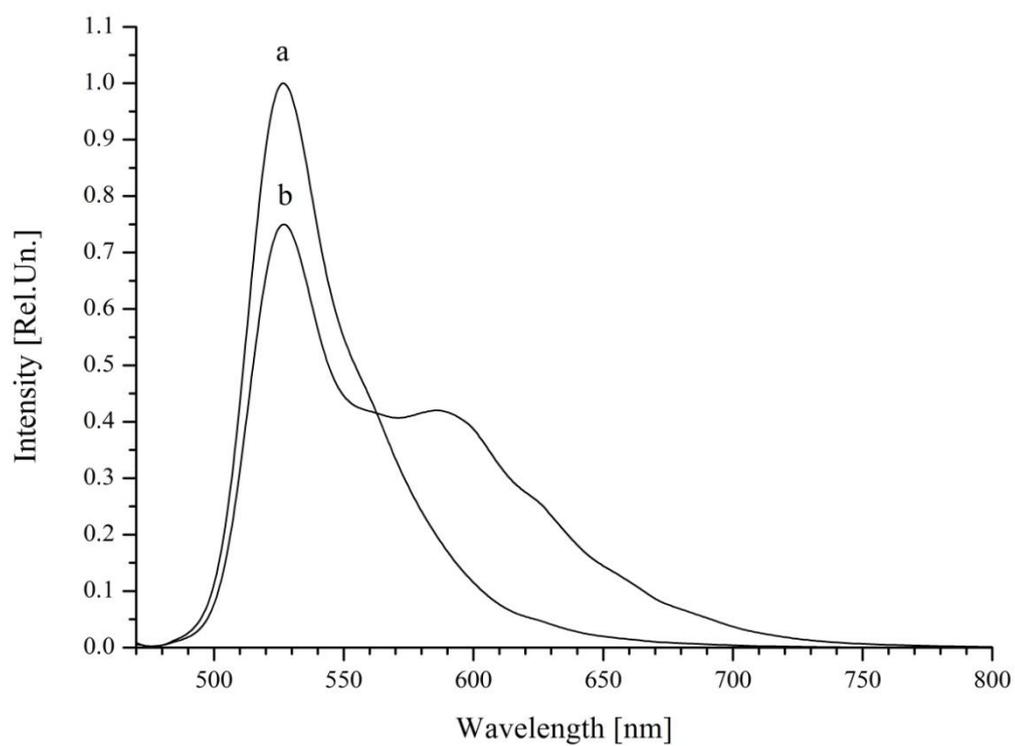

**Fig. 3** Fluorescence spectra of binary and ternary AO complexes: (a) AO-DNA and (b)AO-DNA-EB. [DNA]-1.4·$10^{-4}$ M (P), [AO]-0.14·$10^{-4}$ M, [EB]-0.14·$10^{-4}$ M, [NaNO$_3$]-$10^{-2}$ M.

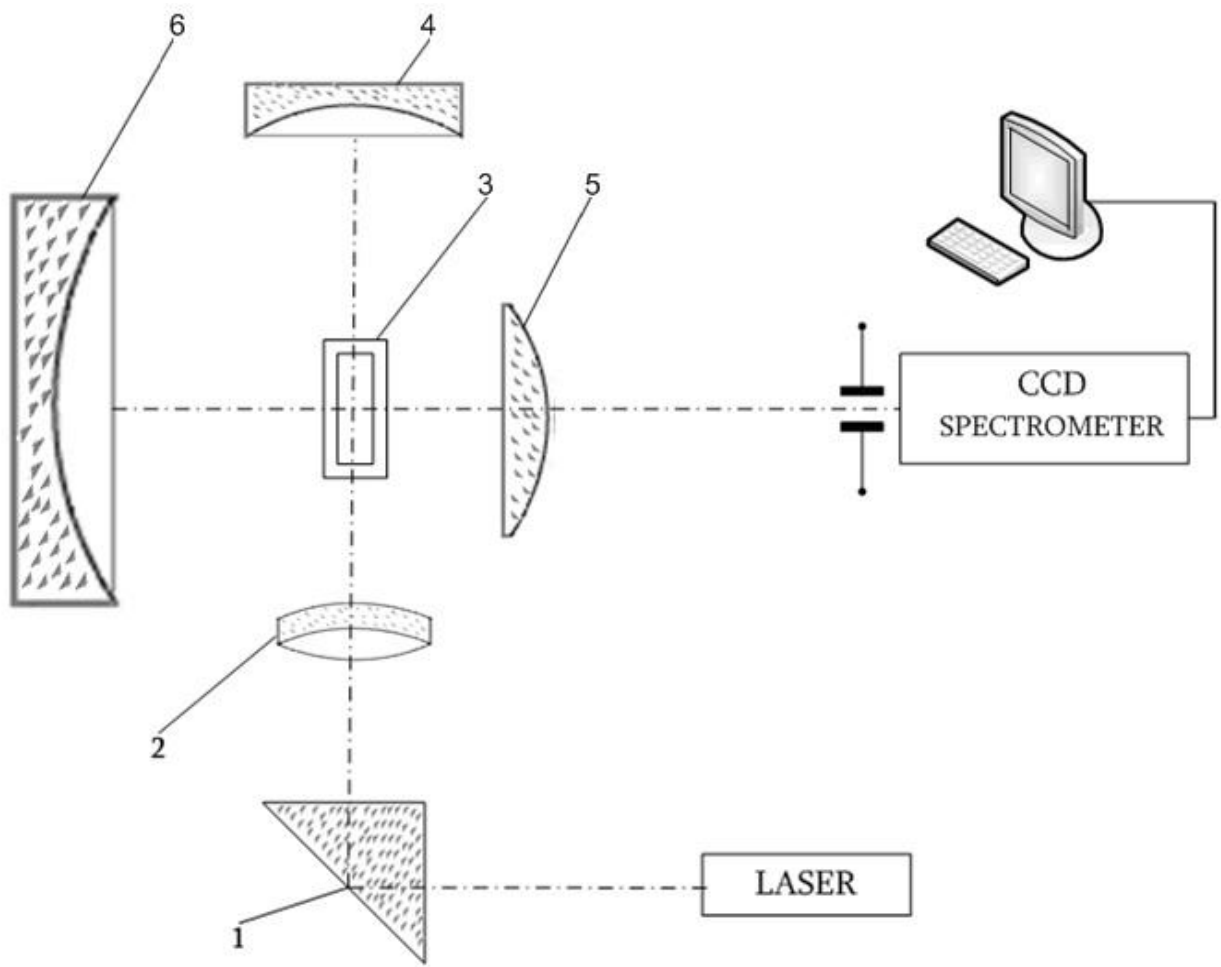

**Fig.4** Lighting system applied for laser induced fluorescence registration in rectangular cells (40 and 80µl). In this case, the laser beam passes through the turning prism 1 and gets on focusing lens 2 with a focal length of 18mm. The focal plane of the lens is coincided with the center of the sample. Exciting beam of radiation passing through the cuvette 3 with sample and a spherical mirror 4 is sent back and refocused at the center of the sample. Thus, the output window of the laser and a spherical mirror 4 forms multi-step system in which light can make a 6-8 passes. The projection system consisting of a lens 5 can form an image of the sample on the entrance slit of the CCD spectrometer. Mirror 6 collects light scattered away from the sell, and focuses it into the center of the cuvette.

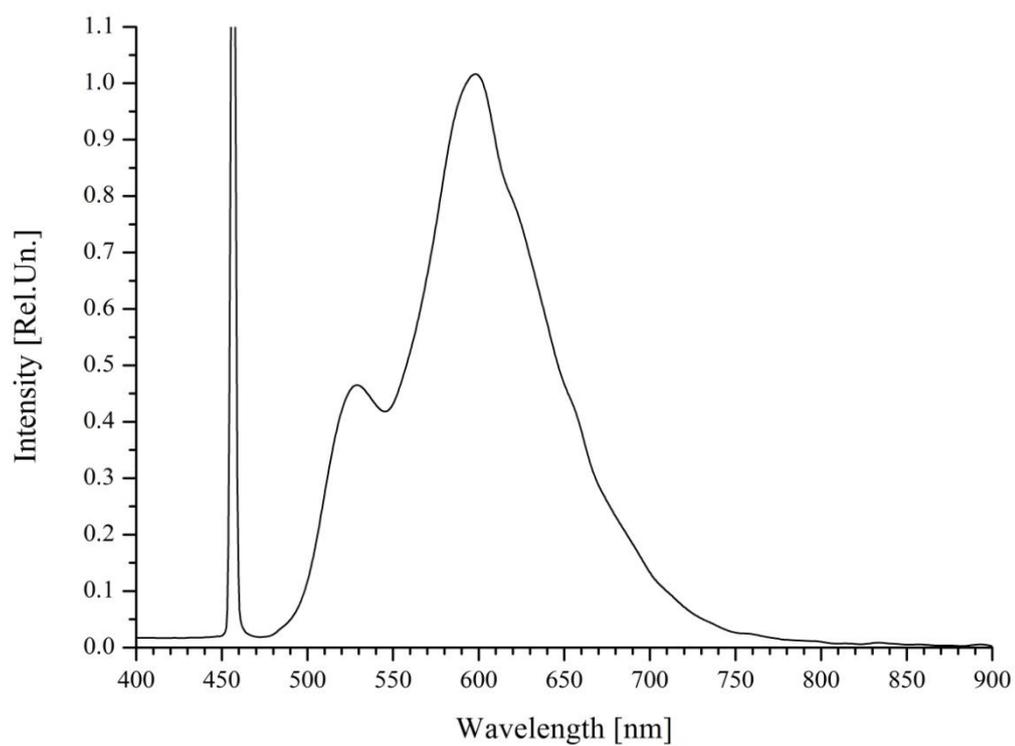

**Fig. 5** Fluorescence spectra of ternary AO-DNA-EB complexes. [DNA]-$1.4 \cdot 10^{-4}$ M (P), [AO]-$0.14 \cdot 10^{-4}$ M, [EB]-$0.14 \cdot 10^{-4}$ M, [NaNO$_3$]-$10^{-2}$ M.

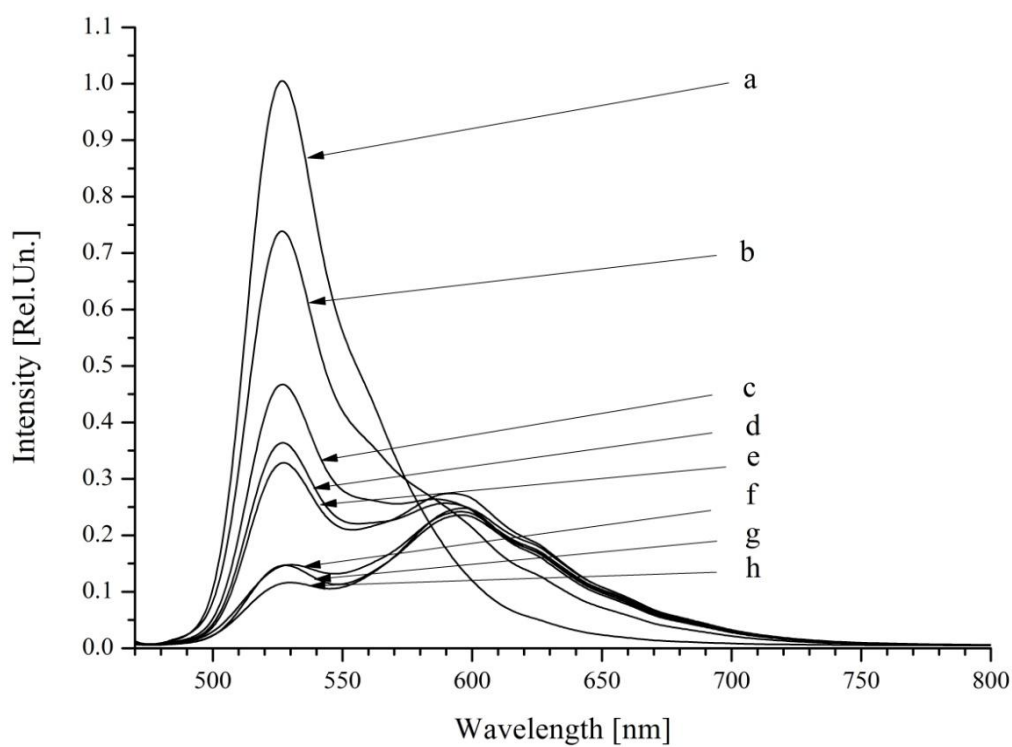

**Fig.6** Fluorescence spectra of double and ternary DNA-AO and DNA-AO-EB complexes where concentration of DNA changes the distance between donor AO and acceptor EB. (a) DNA-AO [DNA]-2.8·$10^{-4}$ M (P), [AO]-0.14·$10^{-4}$, [EB]-0.14·$10^{-4}$ M, [NaNO$_3$]-$10^{-2}$ M. (b) DNA-AO-EB [DNA]-10·$10^{-4}$; (c) DNA-AO-EB [DNA]-8·$10^{-4}$; (d) DNA-AO-EB [DNA]-6·$10^{-4}$; (e) DNA-AO-EB [DNA]-4·$10^{-4}$; (f) DNA-AO-EB [DNA]-2.8·$10^{-4}$; (g) DNA-AO-EB [DNA]-1.4·$10^{-4}$; (h) DNA-AO-EB [DNA]-$10^{-4}$.

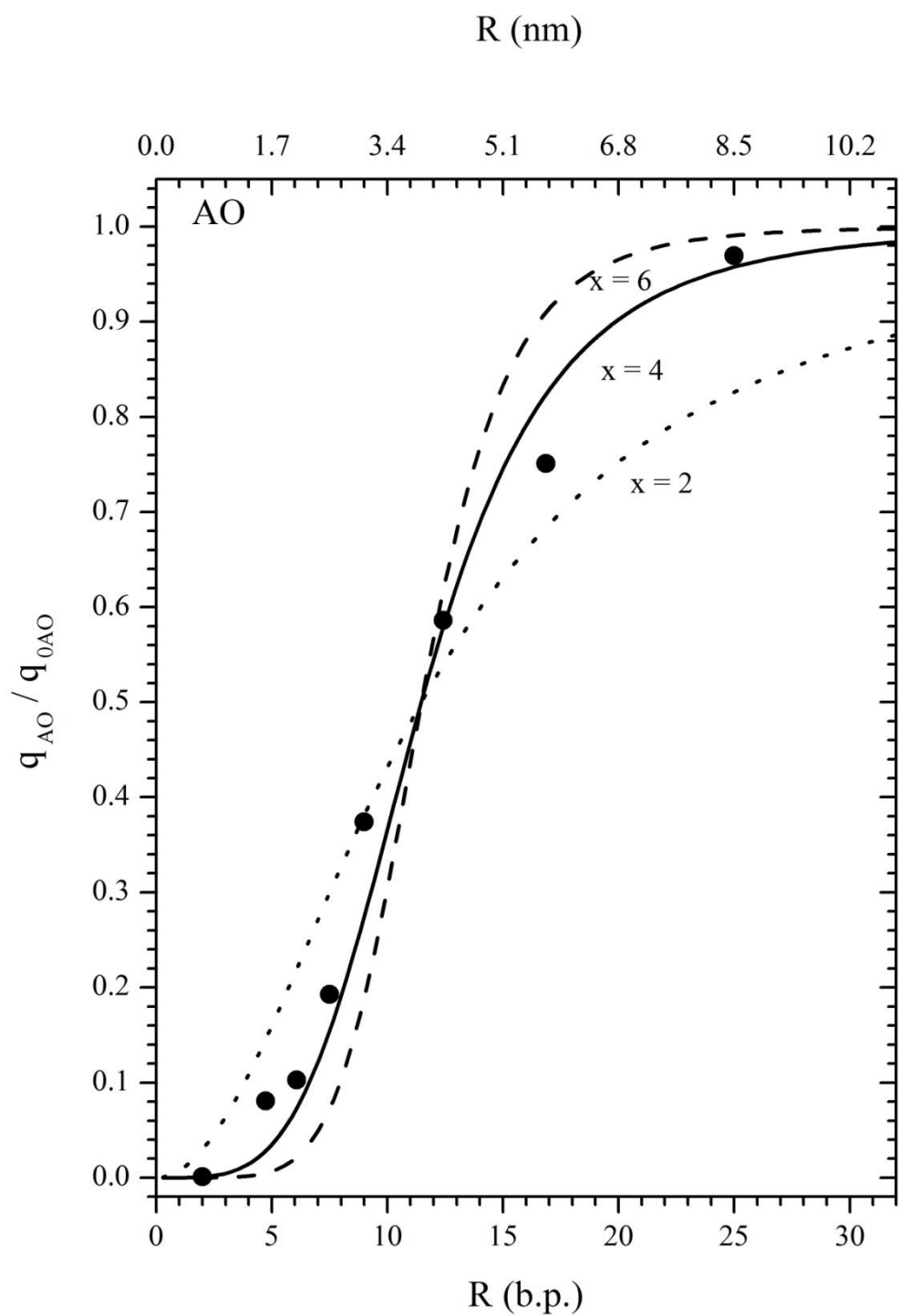

**Fig. 7** Ignition curves of AO fluorescence depending on the distance between AO and EB intercalated in DNA.

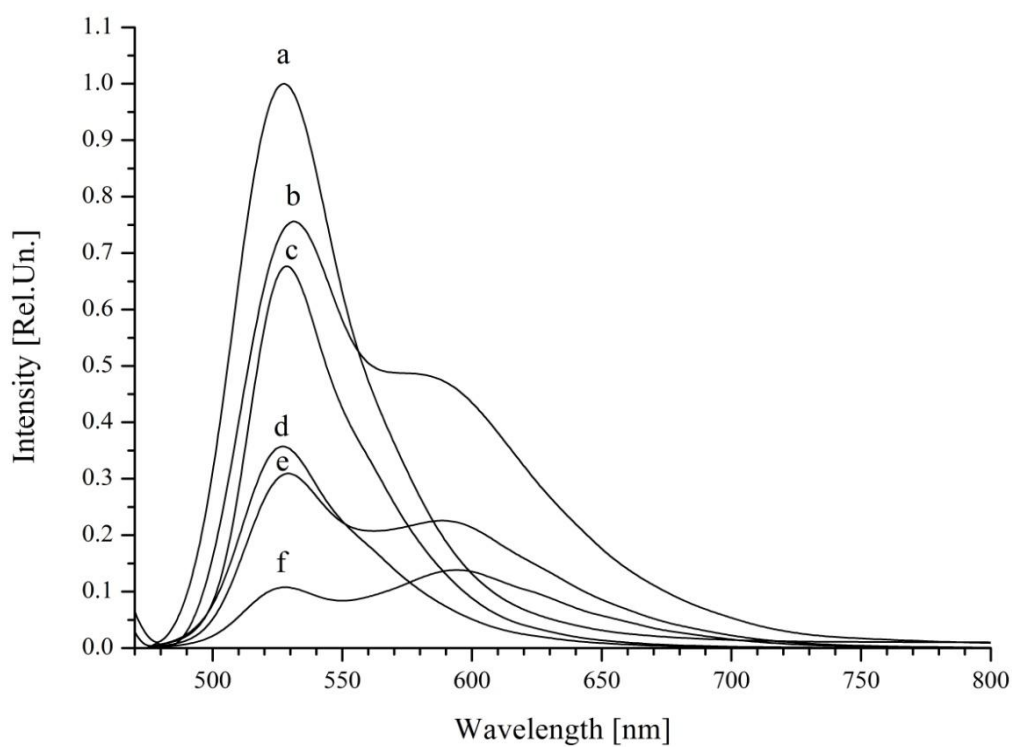

**Figs. 8** Influence of Cu(II) and laser irradiation ($\lambda$ = 457 nm) on electron excitation energy transfer effectiveness from AO to EB intercalated in DNA. (a) DNA-AO; (b) DNA-AO-EB; (c) DNA-AO-Cu(II); (d) DNA-AO-Cu(II) 10 min irradiation; (e) DNA-AO-Cu(II)-EB; (f) DNA-AO-Cu(II) 10 min irradiation +EB. [DNA]-9.6·$10^{-4}$ M (P), [AO]-0.14·$10^{-4}$, [EB]-0.14·$10^{-4}$ M, [CuCl$_2$]0.14·$10^{-4}$ M, [NaNO$_3$]-$10^{-2}$ M.

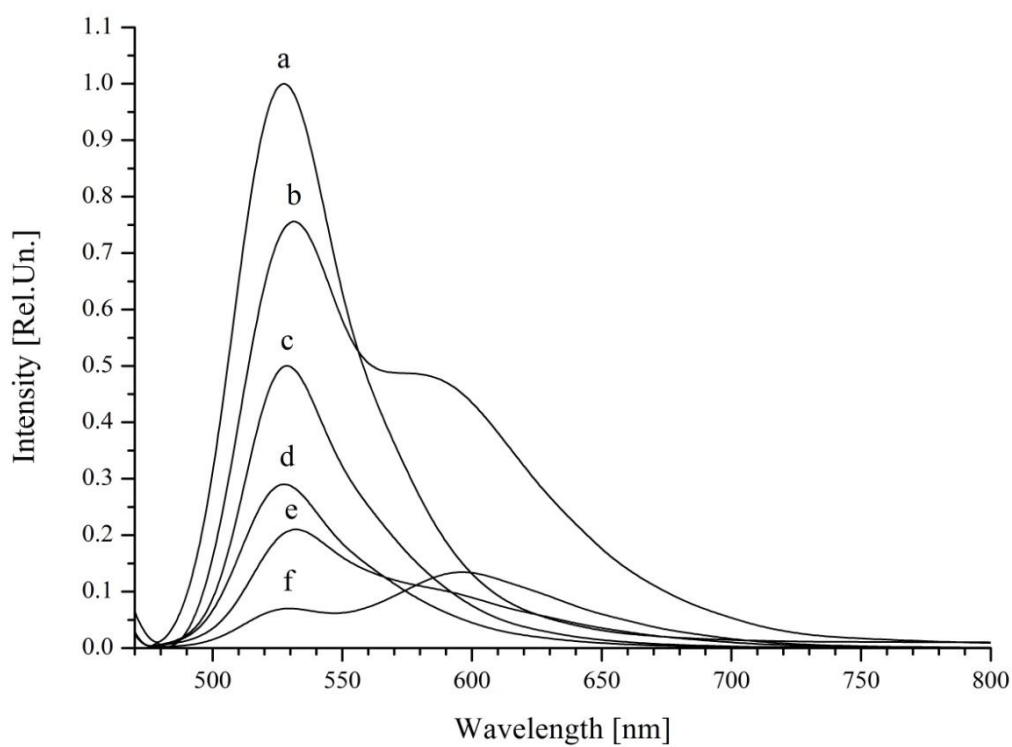

**Figs. 9** Influence of Cu(I) ([ascorbic acid]/[Cu$^{2+}$] 2:1) and laser irradiation ($\lambda$ = 457 nm) on electron excitation energy transfer effectiveness from AO to EB intercalated in DNA. (a) DNA-AO; (b) DNA-AO-EB; (c) DNA-AO-Cu(I); (d) DNA-AO-Cu(I) 10 min irradiation; (e) DNA-AO-Cu(I)-EB; (f) DNA-AO-Cu(II) 10 min irradiation +EB. [DNA]-9.6·10$^{-4}$ M (P), [AO]-0.14·10$^{-4}$, [EB]-0.14·10$^{-4}$ M, [CuCl$_2$]-0.14·10$^{-4}$ M, [AA]-0.24·10$^{-4}$ M, [NaNO$_3$]-10$^{-2}$ M.

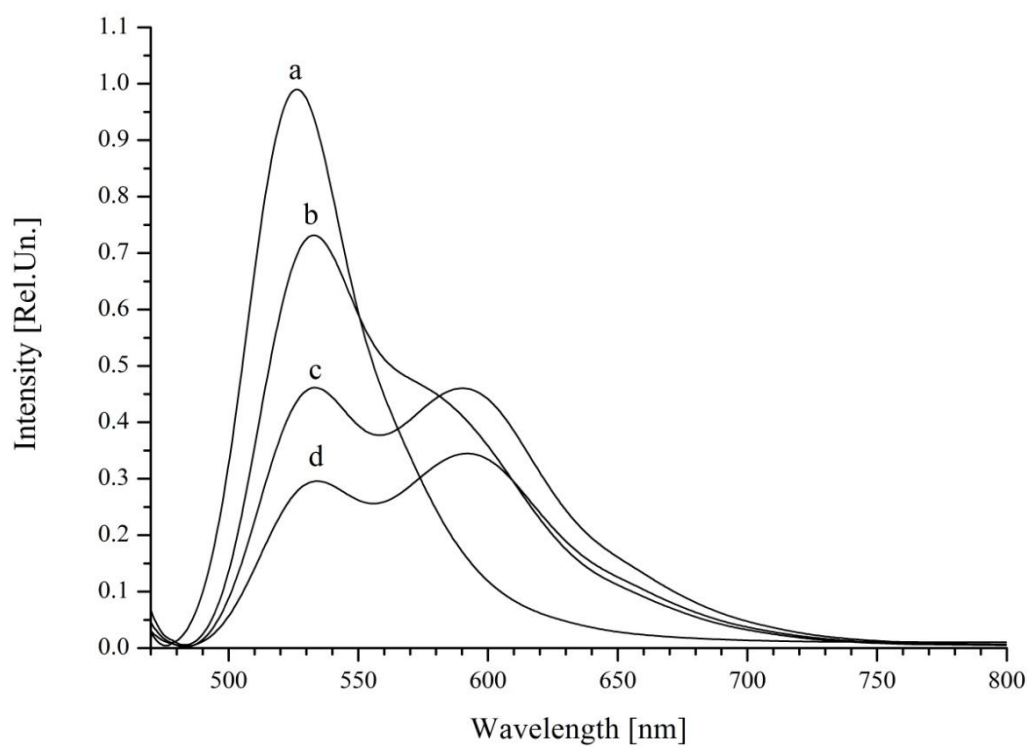

**Fig. 10** Quenching of fluorescence by Ag[+] ion in DNA-AO-EB complex. (a) DNA-AO, (b) DNA-AO-EB- Ag[+] (C$_1$), (c) DNA-AO-EB-Ag[+] (C$_2$), (d) DNA-AO-EB-Ag[+] (C$_3$). [DNA]-2.8·10[-4] M (P), [AO]-0.14·10[-4] M, [EB]-0.14·10[-4] M, [Ag[+]]-0 (C$_1$), [Ag[+]]-6.0·10[-6] M (C$_2$), [Ag[+]]-30.0·10[-6] M (C$_3$), [NaNO$_3$]-10[-2] M. $\lambda$ =460 nm

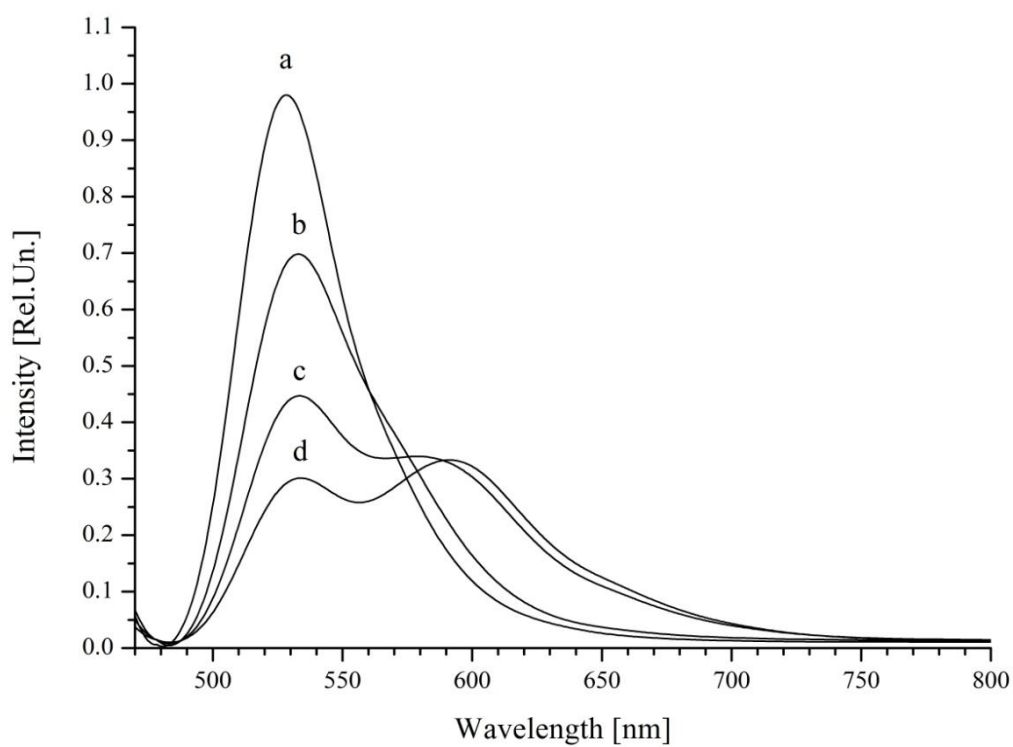

**Fig. 11** Quenching of fluorescence by AgNPs in DNA-AO-EB complex. (a) DNA-AO, (b) DNA-AO-EB- AgNPs ($C_1$), (c) DNA-AO-EB-AgNPs ($C_2$), (d) DNA-AO-EB-AgNPs ($C_3$). [DNA]-$2.8 \cdot 10^{-4}$ M (P), [AO]-$0.14 \cdot 10^{-4}$ M, [EB]-$0.14 \cdot 10^{-4}$ M, [AgNPs]-0 ($C_1$), [AgNPs]-$6.0 \cdot 10^{-6}$ M ($C_2$), [AgNPs]-$18.0 \cdot 10^{-6}$ M ($C_3$), [NaNO$_3$]-$10^{-2}$M. $\lambda$ =460 nm.

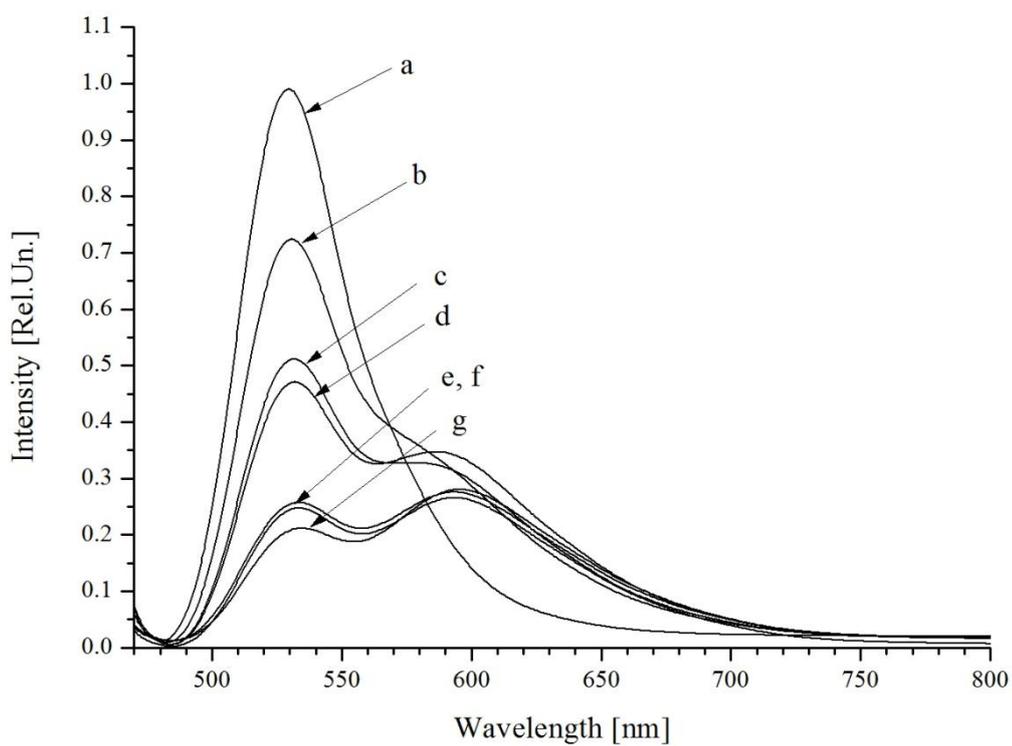

**Fig.12** Heating effect on DNA solution located in a hermetic test-tube in thermostat at various temperatures t= 50, 60, 70, 80 and 90 $^o$ C. (a) DNA-AO, (b) DNA-AO-EB, (c) DNA-AO-EB 50$^o$C, (d) DNA-AO-EB 60$^o$C, (e) DNA-AO-EB 70$^o$C, (f) DNA-AO-EB 80$^o$C, (g) DNA-AO-EB 90$^o$C. [DNA]-2.8·$10^{-4}$ M (P), [AO]-0.14·$10^{-4}$ M, [EB]-0.14·$10^{-4}$ M, [NaNO$_3$]-$10^{-2}$ M.

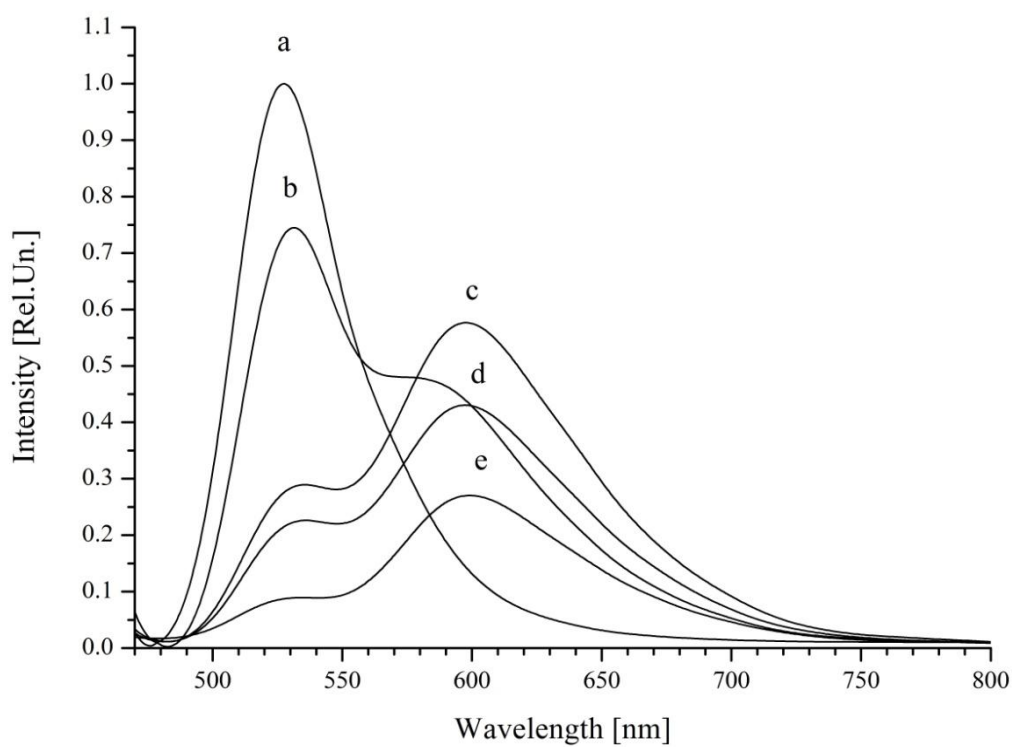

**Fig.13** Fluorescence spectra for ternary complexes AO-DNA-EB. The complexes were prepared as follows: DNA solutions put in hermetic test-tubes were kept in vitro in thermal bath at the temperature $100^0$ C for different periods of time (5, 10 and 20 min). (a) DNA-AO, (b) DNA-AO-EB (0 min), (c) DNA-AO-EB (5 min), (d) DNA-AO-EB (10min), (e) DNA-AO-EB (20min). [DNA]-$2.8 \cdot 10^{-4}$ M (P), [AO]-$0.14 \cdot 10^{-4}$ M, [EB]-$0.14 \cdot 10^{-4}$ M, [NaNO$_3$]-$10^{-2}$ M.